\documentstyle[12pt,aaspp4]{article}
\lefthead{Germany et al.}
\righthead{SN 1997cy}

\begin{document}

\title{SN1997cy/GRB970514 --- A New Piece in the GRB Puzzle?}

\author{Lisa M. Germany; lisa@mso.anu.edu.au}
\affil{Research School of Astronomy and Astrophysics,
  The Australian National University,\\
  Private Bag, Weston Creek P.O., ACT 2611, Australia}

\author{David. J. Reiss; reiss@astro.washington.edu}
\affil{Department of Astronomy, University of Washington, \\
  Box 351580, Seattle, WA 98195-1580}

\author{Elaine. M. Sadler; EMS@Physics.usyd.edu.au}
\affil{School of Physics, University of Sydney, \\
  NSW 2006, Australia}

\author{Brian P. Schmidt; brian@mso.anu.edu.au}
\affil{Research School of Astronomy and Astrophysics,
  The Australian National University,\\
  Private Bag, Weston Creek P.O., ACT 2611, Australia}

\author{C. W. Stubbs; stubbs@astro.washington.edu}
\affil{Department of Astronomy, University of Washington, \\
  Box 351580, Seattle, WA 98195-1580}

\begin{abstract}
 
 We present observations of SN~1997cy, a supernova (SN) discovered as
 part of the Mount Stromlo Abell Cluster SN Search (\cite{reiss98}),
 which does not easily fit into the traditional classification scheme
 for supernovae. This object's extraordinary optical properties and
 coincidence with GRB970514, a short duration gamma ray burst (GRB),
 suggest a second case, after SN~1998bw/GRB980425, for a SN--GRB
 association. SN~1997cy is among the most luminous SNe yet discovered
 ($M_R << -20.3$, $H_0=65$), and has a peculiar spectrum.  We present
 evidence that SN 1997cy ejected approximately 2.6$M_\odot$ of
 $^{56}$Ni, supported by its late-time light curve, and FeII/[FeIII]
 lines in its spectrum, although it is possible that both these
 observations can be explained via circumstellar interaction. While
 SN~1998bw and SN~1997cy appear to be very different objects with
 respect to both their gamma ray and optical properties, SN~1997cy and
 the optical transient (OT) associated with GRB970508 have roughly
 similar late-time optical behavior.  This similarity may indicate
 that the late-time optical output of these two intrinsically bright
 transient events have a common physical process. Although the
 connection between GRB970514 and SN~1997cy is suggestive, it is not
 conclusive. However, if this association is real, followup of short
 duration GRBs detected with BATSE or HETE2 should reveal objects
 similar to SN1997cy.\end{abstract}


\section{Introduction}
Gamma Ray Bursts (GRBs) are among the most energetic and enigmatic
phenomena studied in astronomy. The discovery of optical transients in
association with these events (\cite{vanp97}; \cite{bond97}), and the
evidence of their cosmological origin (\cite{metzger97};
\cite{kulkarni98}; \cite{djorgovski98}) have demonstrated the enormous
energies involved with these intriguing objects.  However, the
association of GRB980425 with SN~1998bw has turned the subject on its
head.  SN~1998bw, located at z=0.0085 (\cite{tinney98}), is the
nearest GRB yet optically identified by a factor of 100, and is
intrinsically fainter than previous GRB optical transients by several
orders of magnitude. Its photometric and spectral resemblance to a
type Ic SN, objects which are thought to result from the core collapse
of massive stars whose outer envelopes have been stripped away either
by binary interaction or a stellar wind, indicates that at least some
GRBs are associated with this type of event.  However, does this
object represent the same physical situation as the optical
counterparts to the more distant GRBs which are on the order of $10^5$
times brighter?

The coincidence of GRB980425 with a peculiar SN suggests that other
such associations are likely.  Unfortunately, the statistics are not
usually compelling for singling out individual objects due to the
large numbers of GRBs and their poor positional information.
\cite{ww98} have cross-correlated the SN and Burst and Transient
Source Experiment (BATSE) catalogs and have identified a positive
correlation with SN~Ic. However, \cite{kippen98} have used a more
sophisticated error model for the BATSE data and found no correlation.
\cite{wes98} took a different approach and looked for GRB
assosciations for three SNe, SN~1992ar, SN~1997cy and SN~1997ef, which
exhibit the high luminosity and peculiar spectra displayed by
SN~1998bw.  Of these three objects, only SN~1997cy, presents a
compelling case for a second SN/GRB association.

In this paper, we detail the spectroscopic and photometric properties
of SN~1997cy.  We present the optical discovery and follow up
observations in section 2, we provide evidence that GRB980514 is
associated with SN~1997cy in Section 3, and in section 4, we explore
the properties of the GRB/SN.


\section{Photometric Observations}

SN~1997cy was discovered on CCD images taken 1997~July~16
(\cite{iauc97}) as part of the Mount Stromlo Abell Cluster Supernova
Search (\cite{reiss98}).  The SN occurred in an anonymous, faint
($M_V$=-17.7 within a 8 kpc radius, $H_0=65$~km~s$^{-1}$~Mpc$^{-1}$),
blue $(V-R=0.1\pm0.1)$, low surface brightness galaxy near Abell
Cluster 3266 (\cite{abell89}) at a redshift of $z=0.063$ (cluster
redshift $z=0.059$; \cite{teague90}).  The SN's position
at $\alpha=4^h32^m54.81^s$,
$\delta=-61^{\circ}42^{\prime}57.9^{\prime\prime}$ (J$2000$) is offset
from the centre of its host galaxy by $0.9^{\prime\prime}$ E and
$1.4^{\prime\prime}$ N.  Unfortunately, the discovery image was the
first of the observing season and the most recent pre-discovery image,
where the SN is {\it not} present, was taken $\sim$4 months earlier on
1997~March~12.  Therefore the date of explosion is not well
constrained.

The first spectrum of SN~1997cy was obtained on 1997~July~24 with the
Danish 1.5m telescope in the wavelength range 3200 - 8800 \AA\ at a
resolution of 20 \AA\ (\cite{benetti97}).  We obtained two further
spectra on 1997~August~9 and 1998~June~26 with the Mount Stromlo and
Siding Spring Observatory's (MSSSO) 2.3m telescope.  These later
spectra encompass a wavelength range of 3900 - 7500 \AA\ at a
resolution of 4.4 \AA, were wavelength calibrated using observations
of a Cu-He lamp taken at the position of the SN, and flux calibrated
using sensitivity curves determined from the observation of southern
flux standards chosen from the list of \cite{bessellPASP}.

We monitored SN~1997cy photometrically in the MACHO $V_M$ and $R_M$
passbands (\cite{stubbs93}; Bessell \& Germany 1999) on the MSSSO
1.3m telescope, and in the $BV(RI)_{KC}$ system with the MSSSO 2.3m
telescope for 15 months. Additional photometric and spectroscopic
observations have been taken at European Southern Observatory and will
be published separately (\cite{turatto99}).

In order to properly determine a SN magnitude from a CCD frame, one
must first remove the light from the host galaxy.  We accomplish this
by subtracting from each observation a template image where the SN is
not present.  We construct template images for the data taken with the
MSSSO 1.3m telescope from $V_M$ and $R_M$ images obtained in good
seeing conditions ($1.4^{\prime\prime}$) on 1997~January~2.  Using the
techniques described in Reiss et al. (1998), we register each of the
images containing the SN to the appropriate template image, convolve
the point spread function (PSF) of the stars in the template with a
kernel to match the PSF of the stars in the registered images, then
match the flux in the registered images and the template.  We then
subtract this intensity-transformed, convolved template from each
registered observation.  Once we have removed the light of the host
galaxy, we calculate the relative photometry between the SN and a
number of local comparison stars through PSF fitting with DoPHOT
(\cite{schechter93}).  This process is detailed in \cite{schmidt98}.

We can not apply the above techniques for extracting SN magnitudes to
the $BV(RI)_{KC}$ MSSSO 2.3m data, because SN~1997cy is still visible
and no template yet exists in these bandpasses.  However, it is
possible to obtain photometry for these images at early times since
the SN is so much brighter than its host galaxy. To measure the
magnitude of the SN, we use the IRAF task PHOT to obtain differential
photometry, within a $3^{\prime\prime}$ aperture, between the SN and
the local comparison stars, and apply a correction for the galaxy
brightness. We calculate the galaxy-correction for each filter
(ranging in size from 0.2 - 0.4 mag for $B$, 0.1 - 0.3 mag for $V$,
and 0.1 - 0.2 mag for $R_{KC}$ and $I_{KC}$) based on the galaxy
magnitude. This magnitude is derived from the MSSSO 1.3m $V_M$ and
$R_M$ template images within this same $3^{\prime\prime}$ aperture and
corrected to $BV(RI)_{KC}$ using the $V_M$ and $R_M$ transformations
calculated by \cite{bessell98}.

We put the local comparison stars onto the $BV(RI)_{KC}$ standard system
under photometric conditions obtained with the Cerro-Tololo Interamerican
Observatory (CTIO) 1.5m on 1999~February~18 and 1999~March~9 and the CTIO
0.9m on 1999~February~28. We determined the transformation coefficients to
the $BV(RI)_{KC}$ standard system and the airmass coefficients for each
night from observations of multiple Landolt fields (\cite{landolt92}) taken
over a wide range of airmasses.  We measured PSF magnitudes for the standard
stars and local comparison stars using DAOPHOT (Stetson 1987) and give the
standard magnitudes for the comparison stars in Table~\ref{tbl-1}.  The
statistical uncertainties in the measurements are included in parentheses.


We cannot calibrate the local comparison stars in the $V_M$ and $R_M$
system in the same way since no record is kept as to whether or not
conditions are photometric.  However, \cite{bessell98} have determined
transformation equations to convert the magnitudes of normal giant stars in
the $BV(RI)_{KC}$ standard system to the $V_M$ and $R_M$ system.  These
transformations are:
\begin{equation}
  V_M = 0.153(B-V) + V
\end{equation}
\begin{equation}
  R_M = -0.154(V-I) + R
\end{equation}

We apply Equations 1 and 2 to the photometry of the local comparison stars
in the $BV(RI)_{KC}$ system to calculate their magnitudes in the $V_M$ and
$R_M$ system.  We then determine the magnitude of the SN in each of the
images from the MSSSO 1.3m from the relative photometry between the SN and
the calibrated local standards shown in Figure~\ref{fig1} and
Table~\ref{tbl-1}.

SN~1997cy is at $z=0.063$ and k-corrections are not negligible for
this object.  Unfortunately, the peculiar nature of SN~1997cy
precludes the use of other SNe for calculating k-corrections, and we
have based our corrections on the spectra presented here, which have
been extrapolated as necessary. Since the corrections for $V_M$ to
$V^{(z=0)}$ and $R_M$ to $R_{KC}^{(z=0)}$ are an order of magnitude
larger than correcting $V$ and $R_{KC}$ to $z=0$, we add constants to
the $V_M$ and $R_M$ data to bring them into alignment with the
k-corrected $V$ and $R_{KC}$ data respectively.  SN~1997cy is a unique
event without good spectrophotometry, and therefore these
k-corrections are uncertain and estimated to be about as good as their
size, $\sim 0.15$ mag.  Due to the extreme luminosity of this object,
however, the uncertainty in the k-corrections has little effect on our
conclusions.  Table~\ref{tbl-2} and Table~\ref{tbl-3} summarize the
photometry and k-corrected photometry obtained for SN~1997cy, and
Figure~\ref{fig2} shows the light curves.  The $V$ and $R_{KC}$ light
curves are well sampled by the $V_M$ and $R_M$ observations taken on
MSSSO~1.3m.


\section{Is SN~1997cy GRB970514?}

From the optical data, it is unclear exactly when SN~1997cy exploded, but it
was certainly sometime between 1997~March~12 and 1997~July~15, based on the
discovery and pre-discovery images. During this 4 month period, 119 BATSE
events were detected (\cite{brainerd98}), two of which overlapped
SN~1997cy's position  within 2$\sigma$.  These were GRB970403
(1.92$\sigma$, 17.3$^\circ$ from SN) and GRB970514 (0.23$\sigma$,
0.88$^\circ$ from SN).  GRB970514's proximity is clearly unexpected --- it
is the closest event to SN~1997cy's position yet detected by BATSE (April
1991 through May 1999) by a factor of 2.5.  Using the prescription of
\cite{ww98}\footnote{ $p = 1-\Pi^N_{i=1}(1-\left(\pi(0.23 \sigma_i)^2 \over
    4\pi\right))$, where $\sigma_i$ is the 1$\sigma$ BATSE error circle for
  each object measured in radians. We note that this approximation assumes
  that none of the error circles overlap --- a reasonable approximation in
  this situation.}, the probability that a 0.23$\sigma$ event will occur
randomly given the 119 BATSE events and their respective error circles, is
1.7\% --- an unlikely occurence.  The fact that two objects are found
within a 2$\sigma$ radius, however, is not surprising.  Over this time-frame
with the 119 events observed, $2\pm 1.5$ random coincidences of this
significance are expected.

GRB970514 had a smaller error circle (3.7$^\circ$) than a typical BATSE
burst, and the above calculation provides the probability of a chance
association between all bursts and SN~1997cy, regardless of their
uncertainty.  Unfortunately, BATSE error circles span a wide range of sizes.
Just three objects contain over 30\% of the probability for chance
association and more than half of the probability of coincidence is
contained in only 10\% of the most poorly determined positions.  With this
approach, if SN~1997cy had been 5$^\circ$ (0.23$\sigma$) away from 
GRB970627 ($21.9^\circ$ error circle), the likelihood of a chance
coincidence would be the same as what was calculated above (1.7\%). A
different approach is to ask, given 119 events, what is the probability that
one will fall within 0.88$^\circ$. This is the method we would use if we had
no error model information, and is equivalent to assigning all 119 BATSE
events error circles of 3.7$^\circ$. A calculation analagous to that above
shows that the probability of a chance 0.88$^\circ$ association of the SN
and 119 BATSE events is 0.7\% (a 5$^\circ$ coincidence would have a
20\% probability with this method).

As a sanity check, we cross-correlated the entire BATSE (up through
1999) catalog with the $677$ SNe discovered from 1991 to 1999
(\cite{Cappellaro99}) and looked for SNe which had a GRB burst closer
than 1$^\circ$ within 121 days of discovery. Apart from SN~1997cy, 5
objects emerge, in agreement with the expected number of chance
coincidences, $4.5\pm2.5$, for this data set. Of these 5 objects with
coincidences, two are SN II (SN~1992Z, SN~1992aw) whose spectra were
not noted as being peculiar, two are normal SN Ia (SN~1996av,
SN~1996bx), and the final object is SN~1993J, whose time of explosion
is well known and does not match that of the correlated GRB. We
conclude that there are no other SN/GRB coincidences as compelling as
SN 1997cy/GRB970514.

We have taken a conservative approach in deriving the above
likelihoods of a chance coincidence between GRB970514 and SN~1997cy that
does not take into account the slow spectral and photometric evolution of
SN~1997cy (which suggest it was more than a month old at discovery). If this
were to be taken into consideration, the chance of a random coincidence
decreases further. However, despite this conservatism, we are still subject
to the the vagaries of {\it a posteriori} statistics because we have had
to specify an interesting SN/GRB angular separation -- after the fact. If, for
example, we had chosen a $1-\sigma$ error circle, rather than the
$0.23-\sigma$ error circle (as was found to be interesting in this case),
the chance for a coincidence between SN 1997cy and a GRB is 15\%.
So although a chance correlation between GRB970514 and SN~1997cy is unlikely
--- about 100 to 1, it is still possible.  We emphasize that we
sought a GRB coincidence initially only for this object given its
extreme nature, and while we believe the association is compelling,
it is certainly not conclusive.  We proceed making the reasonable
assumption that SN~1997cy is indeed GRB970514.


\section{Properties of GRB970514 and SN~1997cy}

\subsection{GRB970514}

GRB970514 was detected on 1997~May~14, with BATSE on board the Compton
Gamma Ray Observatory.  The BATSE burst profile indicated that
GRB970514 was a member of the Fast Rise Exponential Decay subclass of
single peak events and had a duration of $\sim0.2$s.  The total
gamma-ray fluence over all channels was ($4.1\pm1.3$)x$10^{-7}$ ergs
cm$^{-2}$, a factor of 10 less than that of GRB980425.  If we assume
that GRB970514 is at the redshift of SN~1997cy and
$H_0=65$~km~s$^{-1}$~Mpc$^{-1}$), this fluence translates into an
estimated burst energy of $\sim4$x$10^{48}$ ergs.  This is more energy
than GRB980425 released (\cite{galama98}) but $\sim10^4$ times less
than other bursts with measured redshifts.  Emission was detected
above 300keV marking it as a high-energy (HE) burst
(\cite{pendleton97}).  These properties set GRB970514 apart from all
other GRBs which have had an identified optical transient.

Given the association of GRB980425 with SN~1998bw, \cite{bloom98} and
\cite{norris98} have proposed a SN associated subclass of GRBs
(S-GRBs), based on the properties of GRB980425.  These S-GRBs have a
simple, broad burst profile with a rounded maximum, no high energy
(NHE) emission, no long-lived X-ray afterglow and prompt radio
emission.  In contrast, GRB970514 contains HE emission and displays a
cusp rather than a rounded maximum.  Unfortunately there was no
immediate radio\footnote{however, late time observations with the
Australia Telescope Compact Array were obtained and are discussed in
Section 4.4.} or X-ray followup of GRB970514 and no definite
properties that can be assigned to a S-GRB subclass have emerged from
these two objects.


\subsection{Spectra of SN~1997cy}   

Figure~\ref{fig3} shows a spectrum of SN~1997cy obtained with the
MSSSO 2.3m telescope on 1997~August~9, 87 days after GRB970514.  Also
plotted are MSSSO 2.3m spectra of SN~1998bw which bracket the age of
SN~1997cy, and a spectrum of SN~1991T, a type Ia SN approximately 85
days after explosion (\cite{phillips92}).  It is clear from these
spectra that the two objects found to be associated with GRBs are very
different.  SN~1998bw is classified as a type Ic SN (\cite{patat98})
due to the absence of Si, H and He lines which are representative of
type Ia, type II and type Ib SNe respectively.  SN~1997cy appears to be
a hybrid object, containing the H lines present in type II SNe as well
as FeII/[FeIII] lines found in late time spectra of type Ia SNe
(\cite{branch83}). Such an object cannot be classified under the
conventional scheme, and is probably best referred to as a peculiar
type II SN as per Benetti et al. (1997).

The dominant feature in the spectrum of SN~1997cy is an intense
H$\alpha$ emission line with broad ($\sim 3000$~km~s$^{-1}$) and
narrow ($\sim 300$~km~s$^{-1}$) components.  However, it lacks the
P-Cygni profile of H$\alpha$ normally found in type II SNe, and is
reminiscent of the peculiar type IIn SN~1988Z (\cite{raylee91};
\cite{turatto93}). This line profile suggests that a significant
amount of interaction between SN~1997cy and the surrounding
circumstellar material is taking place (\cite{chugai91}).  The other
main features near 4600\AA \ ($\sim 10,500$~km~s$^{-1}$) and 5300\AA \
in SN~1997cy coincide with the position of [FeIII] typical of late
time type Ia spectra.

The origin of the broad absorption feature extending from 5645\AA \ to
6115\AA \ and centred at 5795\AA \ is unclear.  While He(5876) and
Na~D(5893) are possibilities, the absorption seems far too broad and
centered at too low a velocity for this to make sense.  Given the GRB
connection, one might be tempted to consider that this absorption is
H$\alpha$ from a jet pointed towards our line of sight.  However there
is no comparable H$\beta$ absorption.  Another possibility is that
this is not an absorption feature at all, but rather a region without
emission lines.

Figure~\ref{fig4} shows a spectrum of SN~1997cy taken 408 days past
GRB970514.  H$\alpha$ still dominates the spectrum although it appears
narrower than in the spectrum taken 87 days after GRB970514.  This and
the continued presence of the mysterious dip at 5795\AA \ show the
spectrum of SN~1997cy has not evolved significantly in almost a year.


More recently SN~1999E was found to have a spectrum similar to SN~1997cy.
It displays an intense H$\alpha$ emission line, a broad absorption feature
centred near 5970\AA \ and broad undulations similar to SN~1997cy
(\cite{flipper99}; \cite{Jha99}; \cite{Cap99}). In addition, it was
atypically luminous $M_V < -19.4$ ($H_0=65$ ; \cite{Jha99}) and can been
linked to GRB980910 (\cite{TH99}), although without great certainty [There
are no prediscovery images and this SN is separated by 4.8$^\circ$ from the
GRB position - which itself is uncertain by 6.8$^\circ$].


\subsection{Interpreting the Light Curve of SN~1997cy}

We have constructed a L$_{UVOIR}$ light curve by integrating over the
$V_M$, $R_M$ and $BV(RI)_{KC}$ photometry from the data taken
1997~August 9 and 10, before it was k-corrected.  We derived the
effective wavelengths and flux measurements for each filter by
convolving their filter transmission curves in the rest frame of the
SN (\cite{bessell90}; Bessell \& Germany 1999) with the absolute
spectrophotometry of Vega (\cite{dreiling}).  In keeping with the
observed properties of SN, we extended the IR with a Rayleigh Jeans
tail and truncated the UV with a function that declined much faster
than a black body.  This treatment of the UV is to simulate line
blanketing and represents a lower bound for the flux beyond the $B$
filter. The correction is reasonably secure since most of the flux of
the SN is contained within the wavelengths we have observed; the total
flux extrapolated to lie beyond the wavelength region of the
observations is approximately 20\%.  We held this derived correction
fixed (for lack of better information) for all observations and
ascribed a systematic error equal in size to the correction (20\%) in
the L$_{UVOIR}$ light curve.  We note that if SN~1997cy has a large UV
or IR excess, then the derived bolometric luminosity could be
underestimated.

Figure~\ref{fig5} shows the L$_{UVOIR}$ light curve of SN~1997cy compared to
that of SN~1998bw (Woosley et al. 1998) and a simple model where the late
time flux is solely due to the complete thermalization of gamma-rays from
the decay of 2.6 M$_\odot$ of $^{56}$Co.  SN~1997cy was clearly much brighter
than SN~1998bw and initially declined at the rate of $^{56}$Co.  We believe
the large excursion from the $^{56}$Co decay line is due to circumstellar
interaction which has subsequently decayed.  This picture is consistent with
the presence of FeII/[FeIII] lines (the final decay product of $^{56}$Ni) and
H$\alpha$ in the spectrum (\cite{chugai91}). It is also possible that the
SN is being powered only by circumstellar interaction, and the decay rate
is only a coincidence.  \cite{nomoto99} have modelled SN 1997cy's 
light curve under this latter assumption and find they can reproduce this late-time behaviour with an extremely energetic ($8\times 10^{52}$ erg)
SN explosion expanding into dense circumstellar material. A clumpy
circumstellar medium may allow a somewhat smaller explosion energy.
Regardless of the cause, the energy output is large, with the
integral of  $L_{UVOIR}$ from discovery to 450 days later yielding
$1.6\times10^{50}$ ergs.

SN~1997cy is unlike any other SN observed (except 1999E previously 
discussed), and its unusual nature and coincidence with GRB970514
prompts speculation as to whether it is a SN, or rather an
optical transient associated with a GRB (GRBOT).

GRBOTs have been discovered previously (\cite{vanp97}; \cite{bond97}),
but none are short duration events, and only the GRBOT associated with
GRB970508 (\cite{bond97}) has a convincing redshift ($z=0.835$, Bloom
et al. 1998) and a well sampled $R$ light curve from early to late
times (\cite{F99}).  In order to compare SN~1997cy to the GRBOT
associated with GRB970508 we construct an equivalent $R$ light curve
for SN~1997cy at $z=0.835$ by calculating the flux distribution
measured across the $V_M$, $R_M$ and $BV(RI)_{KC}$ passbands (as
described above), and then redshifting this flux distribution from
$z=0.063$ to $z=0.835$.  We scale the flux down by the square of the
luminosity distances ($q_0=0.1$) and by another factor of $(1+z)$ (to
account for the $d \lambda$ change), extrapolating the data to near UV
wavelengths for complete $z=0.835$ $R$ band coverage.  The $z=0.835$
$R$ band magnitude is synthesized from this flux distribution and the
offset between this and the observed $V_M$ magnitude is applied to all
$V_M$ data to produce the $R$ light curve of SN~1997cy as
it would have been observed at $z=0.835$.  The uncertainty in the
$z=0.835$ $R$ light curve is approximately 0.2 mag, and is dominated
the UV flux extrapolation to cover the redshifted $R$ bandpass.

Figure~\ref{fig6} plots the $z=0.835$ $R$ light curve for both
SN~1997cy and the GRBOT970508 (\cite{F99}).  The overlap region
shows that the gross properties of the two
late-time light curves are similar, with SN~1997cy outshining
GRBOT970508 by about 1.7 mags. While it was earlier demonstrated that
SN~1997cy is declining with a rate remarkably consistent with the 77
day half life of $^{56}$Co, a similar analysis of GRBOT970508's
late-time emission is inconclusive; the data are consistent with a 77
day half life, but are too poor to substantiate this hypothesis.  However,
the general agreement between these two light curves suggests that
SN~1997cy and GRBOT970508 may share similar physical properties at late times. 
It may be worthwhile to investigate the late-time light
curves of suitable GRBOTs and see if the radioactive decay of
$^{56}$Co to $^{56}$Fe is responsible for their late-time energy
output.

Because of the different properties between GRB970514 and all other
GRBs with associated OTs, further comparison is dangerous.  We note,
however, that if SN~1997cy is a GRBOT with properties similar to those
already discovered, then the disparity between the intrinsic fluences
of its associated GRB and other GRBs (GRB970508 is 5000 times more
luminous than GRB970514 given their respective redshifts of $z=0.835$
and $z=0.063$) is naturally explained by relativistic beaming, where
distant GRBs are beamed towards us, but GRB970514 had no preferred
orientation. This model demands that the relativistic beaming factor,
$\gamma$, for GRBs is low, $\gamma < 10$. Larger values of $\gamma$
would make the chance of seeing any gamma rays from SN~1997cy
impossibly small. This conclusion is only valid, however, if we assume
that GRB970514 belongs to the same general class of objects which make
up the observed GRBOTs -- a dubious assumption.

The existence of SN~1998bw/GRB980425 complicates issues because it
appears to be a very different object compared to SN~1997cy/GRB970514
with respect to both its GRB and SN properties, and because the
optical properties of SN~1998bw are very different from other
GRBOTs. If SN~1998bw and SN~1997cy are both GRBs, these two events,
coupled with the significant number of GRBOTs identified at $z>0.8$
(\cite{djorgovski98}), provide strong argument for the luminosity
function of GRBs to be either rapidly evolving, bimodal or very broad.
We prefer the latter two explanations since it is difficult to imagine
how an astrophysical event can evolve by a factor of $10^4$ from $z=1$
to $z=0$. Bimodality is a natural consequence of a beaming model, but
the significant difference between the two nearby SN/GRBs argues that
GRBs could be produced by more than one mechanism.  It is therefore
quite possible, or even likely, that an observed bimodality of the
GRB luminosity function could be a consequence of both beaming and a
variety of explosion mechanisms which lead to GRBs.


\subsection{Late Time Radio Observations of SN~1997cy}

Radio continuum observations of the field of SN~1997cy were made with the 6A
(6~km) configuration of the Australia Telescope Compact Array during a
14-hour session on September~15~1998. The telescope observed simultaneously
in two bands, with central frequencies 1.384 GHz (20 cm band) and 2.496 GHz
(13 cm band) and bandwidth 132 MHz. Data reduction was carried out in the
usual way with the AIPS package. The synthesized beam FWHM was about 6.7
arcsec at 20 cm and 3.6 arcsec at 13 cm.

No radio emission from SN~1997cy was detected at either frequency, and the
RMS noise in the final cleaned maps near the position of the SN was
78 $\mu$Jy/beam at 20 cm and 68 $\mu$Jy/beam at 13 cm.  We therefore adopt
3$\sigma$ upper limits to the radio flux density of SN~1997cy of 0.23 mJy at
20 cm and 0.20 mJy at 13 cm.  At the distance of SN~1997cy, this corresponds
to upper limits of 2.3$\times10^{21}$ W/Hz and 2.0$\times10^{21}$ W/Hz for
the radio luminosity at 20 and 13 cm respectively.

Unfortunately, the upper limit of 2.3$\times10^{21}$ W/Hz for the 20\,cm
radio luminosity of SN~1997cy is not very restrictive because the SN was so
distant.  The 20\,cm radio luminosity of SN~1998bw (the most luminous radio
SN yet discovered) reached a maximum of about 5$\times10^{21}$ W/Hz, but
only exceeded 2.3$\times10^{21}$ W/Hz for a period of about forty days
between 40 and 80 days after explosion (\cite{kulkarni98}).  The radio light
curve of the next most luminous radio SN, the type II SN~1988Z
(\cite{vand93}), rose much more slowly, reaching a 20\,cm luminosity of
1.8$\times10^{21}$ W/Hz almost five years after explosion (at which point
the 20\,cm emission was still rising).

Our radio data, taken about 16 months after explosion, cannot rule out the
possibility that SN~1997cy was as powerful as SN~1998bw (where the 20\,cm
peak was reached less than 50 days after explosion) or SN~1988Z (where the
20\,cm flux density was still rising after almost five years).  Another deep
radio observation in 2-3 years time might set some further constraints on
the late-time radio light curve of SN~1997cy, but unless this interesting SN
is an even more radio-luminous version of SN~1988Z we are unlikely ever to
learn much more about its radio properties.


\section{Conclusion}

SN~1997cy is, to our knowledge, the brightest SN yet discovered
[M$_R$=-20.3 ($H_0=65$) at discovery, and certainly much brighter at
maximum light].  Its extreme luminosity and peculiar type II spectrum
suggest the progenitor was a supermassive star that underwent core
collapse.  In addition, SN~1997cy may claim the most prodigous output
of $^{56}$Ni with our simple model indicating that it put out 30 times
more $^{56}$Ni than is produced in typical core collapse SN. If
instead the late-time light curve is soley due to energy input from
circumstellar interaction, SN 1997cy is still a remarkable event, with
an explosion energy considerably larger than $10^{50}$ ergs. In either
case, SN~1997cy is an excellent candidate for a black hole induced
hypernova (\cite{paczynski}; \cite{iwamoto98}), collapsar
(\cite{woosley93}; \cite{woosley}), or possibly a pair production
supernova (\cite{woosley86}), with the latter being the only model
which can easily output as much as 2.6 M$_\odot$ of $^{56}$Ni.

The levelling off of the light curve at late times and later reversion
to the $^{56}$Co decay line is probably due to energy input from
circumstellar interaction which started to diminish in its effect at
approximately 300 days after explosion.  This conclusion is
supported by the the presence of a narrow and broad component to the
H$\alpha$, and the absence of H$\alpha$ absorption
(\cite{chugai91}). The material illuminated by this circumstellar
interaction may pose problems for the clean environment necessary for
some gamma-rays burst models, but provides the possibility that the
X-ray pulse of the SN could be inverse compton scattered off the fast
electrons formed as the first material from the SN and the surrounding
material collide.

The association of SN~1997cy with GRB970514 is suggestive, but not
conclusive.  If this association is real there are two GRBs with
measured redshifts below $z<0.1$ (\cite{tinney98}; Benetti et
al. 1997) ,and many more at $z>0.5$ (\cite{djorgovski98}). Despite the
obvious problems associated with such small numbers and the
non-uniform way in which SNe are discovered, only a bimodal or very
broad luminosity distribution can easily explain the observed
distribution of these GRB redshifts.  Bimodality naturally occurs
within a beaming model, but it could be complicated by the existence
of two or more populations of objects which give rise to GRBs. One
thing is clear, if a non-negligible fraction of GRBs are SNe at
$z<0.1$, these objects should be easily uncovered if BATSE LOCBURST
events with $<2{^\circ}$ error circles are followed up using the new
breed of wide-field imagers on small telescopes.

The search for optical transients associated with GRBs should
intensify over the next few years given the small positional error
circles achievable with BeppoSAX and HETE2, and the large swaths of
sky that instruments such as the MSSSO~1.3m and other small telescopes
can observe. Such optical followup of GRBs is necessary if we are to
discover if SN/GRB associations like SN~1997cy/GRB970514 and
SN~1998bw/GRB980425 are real, or pure chance.  Only with more data can
we learn how SNe fit into the big picture of GRBs.

The authors would like to thank Sebastian Juraszek and Vince McIntyre
for carrying out the ATCA observations, and Ron Ekers and Mark
Wieringa for allowing us to use their unpublished data.  We gratefully
acknowledge helpful discussions with Marc Kippen on the BATSE data,
Stan Woosley, Ron Eastman, Peter Garnavich, and Kenichi Nomoto on the
light curves and spectra of SN~1997cy, Geoff Bicknell on the formation
of gamma rays, and David Branch on Fe II lines in SN Ia.  We would
also like to thank Professor Jeremy Mould for his generous allocation
of Director's Discretionary time for the Mount Stromlo Abell Cluster
Supernova Search on the MSSSO~1.3m telescope.  D. Reiss and C. Stubbs
gratefully acknowledge the support of the Packard Foundation and the
National Science Foundation.


\clearpage 

\figcaption[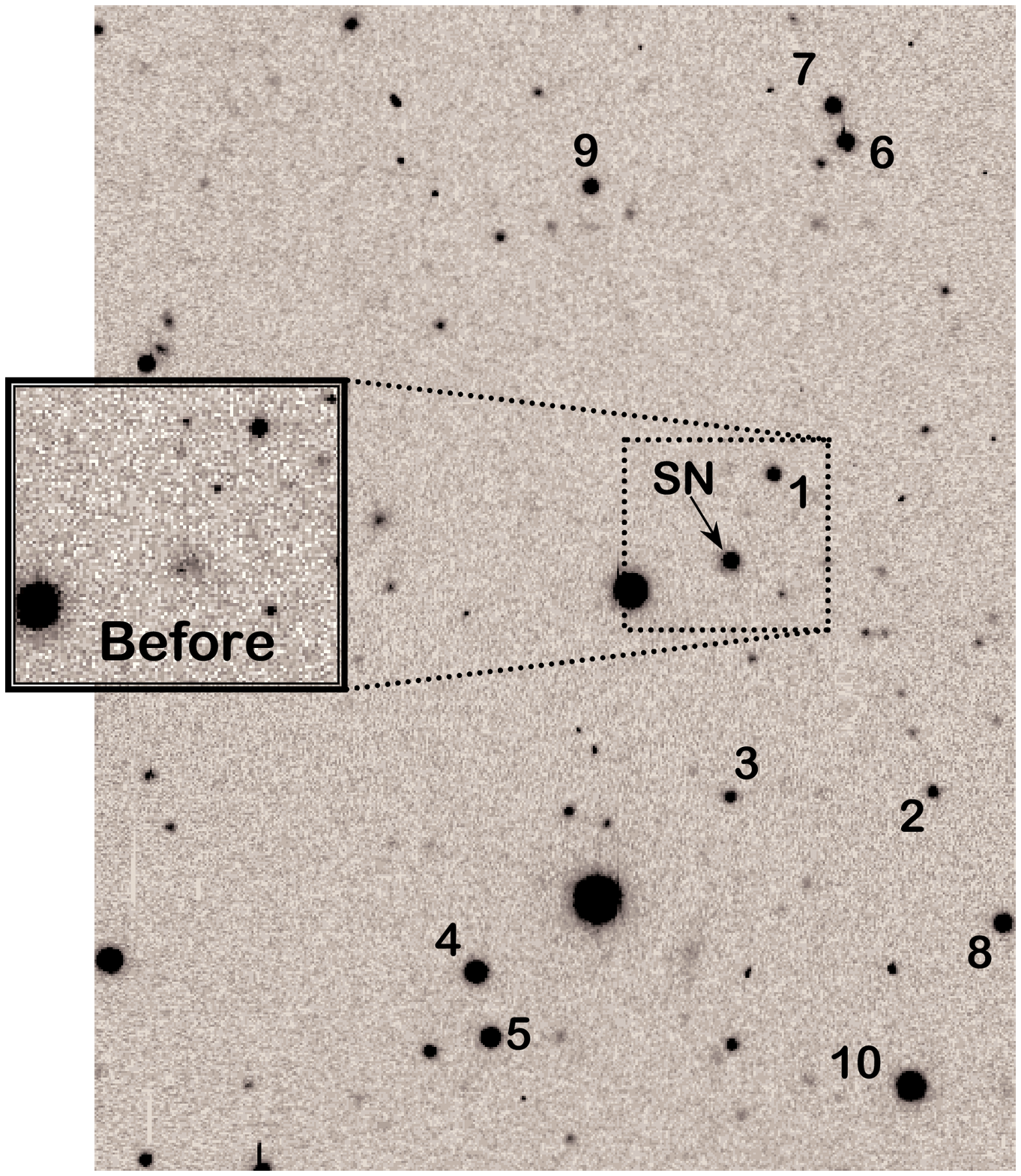]{The field of SN~1997cy showing pre and post SN
  images of the galaxy and the positions of the field reference stars.
  \label{fig1}}

\figcaption[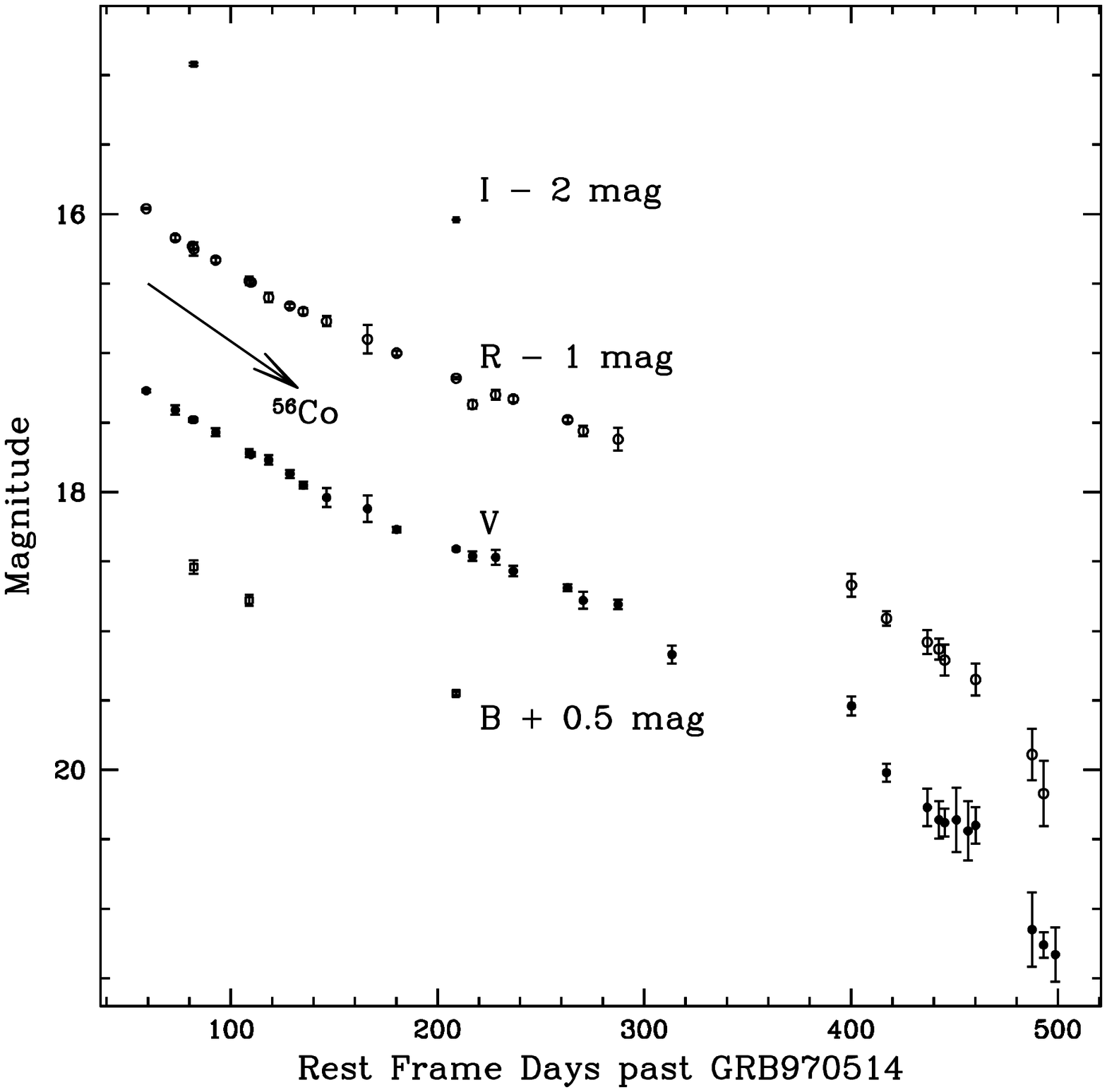]{$BV(RI)_{KC}$ light curves of SN~1997cy shown
  with the decay rate of $^{56}$Co, k-corrected to
  $z=0$.\label{fig2}}

\figcaption[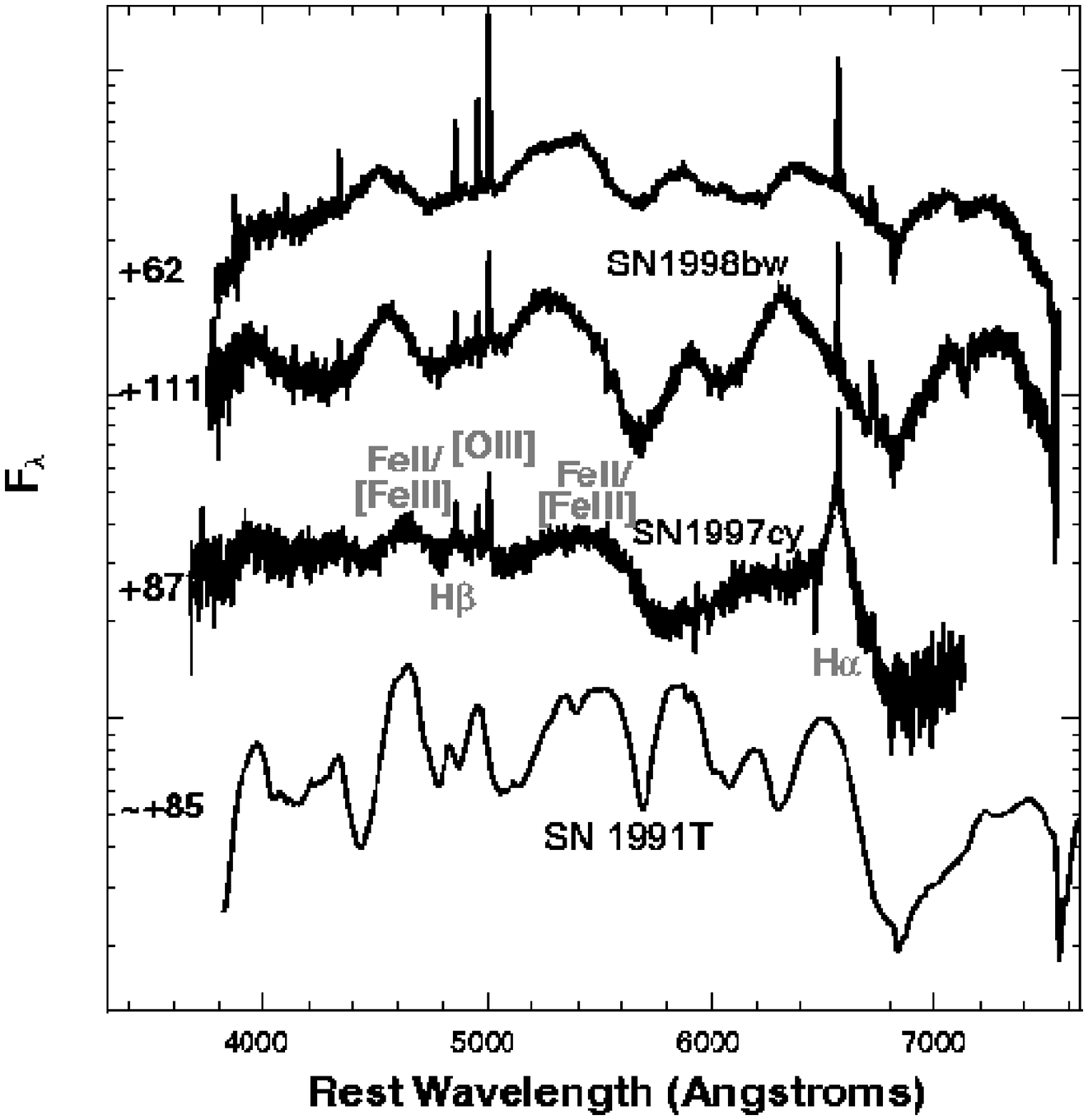]{Spectra of SN~1998bw taken 62 and 111 days
  after GRB980425, SN~1997cy taken 87 days after GRB970514 and
  SN~1991T at 85 days after explosion. SN 1997cy is a distinctly
  different object, but seems to have FeII/[FeIII] features in common
  with SN 1991T at a similar age, and a H$\alpha$ profile similar to
  those SNe thought to be undergoing circumstellar
  interaction.\label{fig3}}

\figcaption[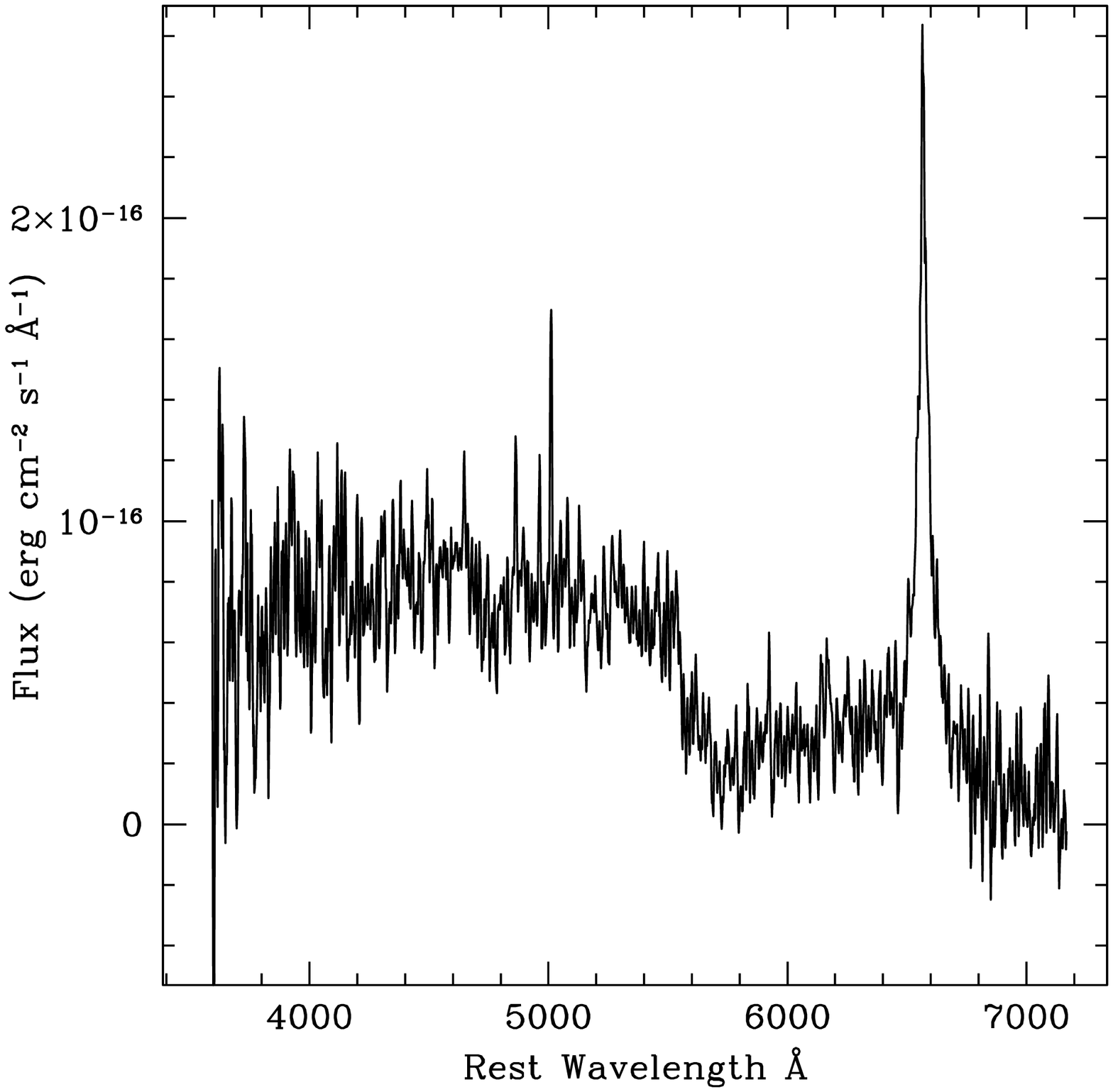]{A binned spectrum of SN~1997cy taken 408 days
  after GRB980514.  The spectrum shows that the SN has evolved little
  in the year since its discovery. \label{fig4}}

\figcaption[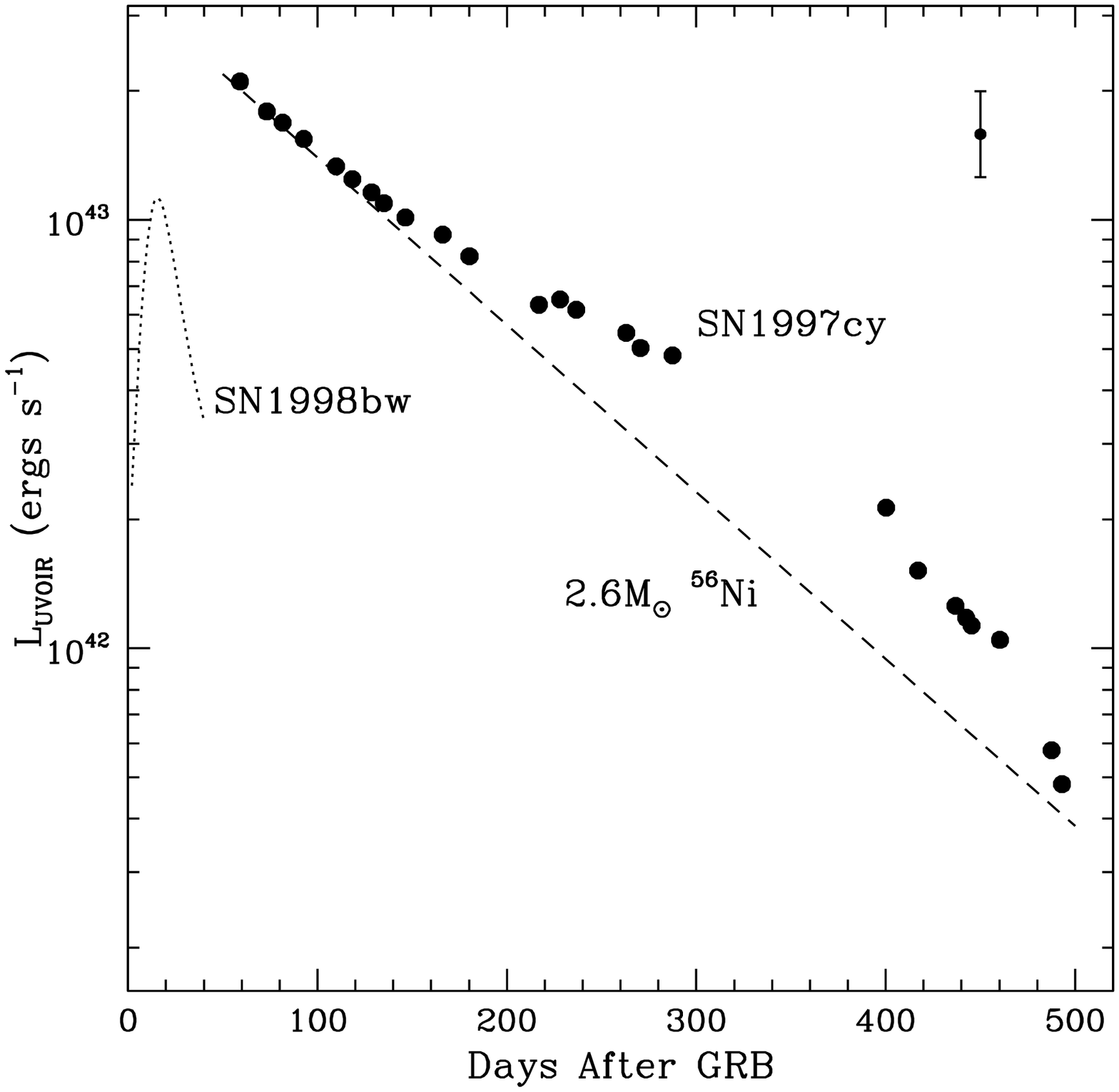]{Bolometric light curve of SN~1997cy compared
  to that of SN~1998bw and a simple model showing the production of 2.6
  M$_\odot$ of $^{56}$Co. The error bar in the top corner represents a
  systematic error of 20\%.\label{fig5}}

\figcaption[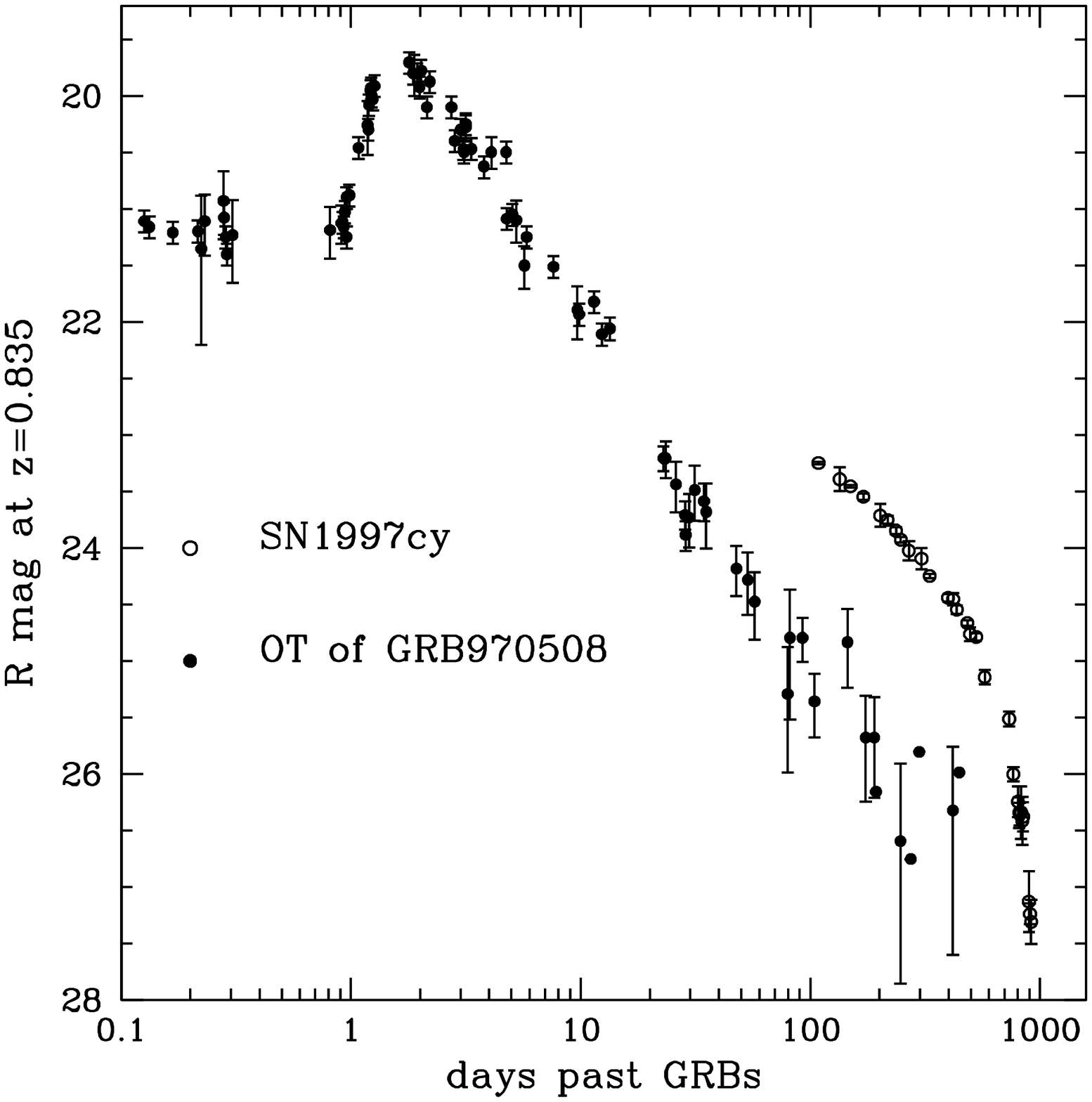]{The $R$ light curve of the OT associated with
  GRB970508 compared to the $R$ light curve of SN~1997cy as it would
  appear at this same redshift.  SN 1997cy is a factor of 3 brighter
  than this GRBOT970508, and exhibits a similar light curve behavior
  in the areas of overlap.\label{fig6}}

\clearpage
\begin{deluxetable}{cccccccc}
\footnotesize
\tablecaption{Calibrated Magnitudes of Field Reference Stars \label{tbl-1}}
\tablewidth{0pt}
\tablehead{
\colhead{Star} & \colhead{B} & \colhead{V} & \colhead{R} & \colhead{I}
}
\startdata
1  & 18.387(0.017) & 17.907(0.011) & 17.607(0.017) & 17.275(0.024)\nl  
2  & 19.506(0.030) & 18.406(0.012) & 17.747(0.019) & 17.151(0.028)\nl  
3  & 18.943(0.188) & 18.440(0.010) & 18.094(0.016) & 17.770(0.019)\nl  
4  & 16.819(0.130) & 15.770(0.009) & 15.156(0.013) & 14.637(0.015)\nl  
5  & 16.932(0.007) & 16.355(0.004) & 16.005(0.007) & 15.669(0.008)\nl  
6  & 18.084(0.025) & 16.932(0.006) & 16.210(0.009) & 15.587(0.010)\nl  
7  & 17.992(0.163) & 17.089(0.009) & 16.519(0.013) & 16.018(0.014)\nl  
8  & 17.503(0.007) & 16.647(0.005) & 16.157(0.007) & 15.680(0.013)\nl  
9  & 18.274(0.124) & 17.513(0.007) & 17.080(0.012) & 16.672(0.013)\nl  
10 & 15.544(0.014) & 14.710(0.006) & 14.231(0.009) & 13.778(0.017)\nl   
\enddata
\end{deluxetable}

\clearpage
\begin{deluxetable}{ccccccccc}
\scriptsize
\tablecaption{Journal of the Observations. \label{tbl-2}}
\tablewidth{0pt}
\tablehead{
\colhead{JD} & \colhead{DATE} &Telescope  & \colhead{B}   & \colhead{$V_M$} & 
\colhead{V}  & \colhead{R} & \colhead{$R_M$} & \colhead{I}\\
\colhead{(2450000+)}
}
\startdata
646.31 &16-07-97&1.3m & &     17.364(0.011) & & &                 16.761(0.005)\nl
661.28 &31-07-97&1.3m & &     17.507(0.106) & & &                 16.968(0.017)\nl
670.21 &09-08-97&1.3m & &     17.570(0.013) & & &                 17.034(0.017)\nl
670.80 &10-08-97&2.3m &18.007(0.05)   & &  17.375(0.01) & 17.104(0.05)       & &    16.911(0.01)\nl 
681.74 &21-08-97&1.3m & &      17.663(0.029) & & &                 17.133(0.019)\nl
698.79 &07-09-97&2.3m &18.210(0.04) &&     17.593(0.03) & 17.321(0.03)\nl
700.22 &08-09-97&1.3m & &      17.828(0.100) & & &                 17.288(0.021)\nl
709.27 &17-09-97&1.3m & &      17.867(0.034) & & &                 17.396(0.034)\nl
720.16 &28-09-97&1.3m & &      17.963(0.029) & & &                 17.457(0.014)\nl
727.05 &05-10-97&1.3m & &      18.044(0.024) & & &                 17.501(0.023)\nl
739.06 &17-10-97&1.3m & &      18.139(0.085) & & &                 17.574(0.036)\nl
760.09 &07-11-97&1.3m & &      18.210(0.095) & & &                 17.703(0.101)\nl
775.06 &22-11-97&1.3m & &      18.363(0.020) & & &                 17.802(0.016)\nl
805.66 &23-12-97&2.3m &18.755(0.01) &&     18.193(0.01) & 17.950(0.01)       &&     17.917(0.01)\nl
814.07 &31-12-97&1.3m & &      18.557(0.034) & & &                 18.172(0.029)\nl
825.96 &12-01-98&1.3m & &      18.569(0.055) & & &                 18.100(0.036)\nl
835.04 &21-01-98&1.3m & &      18.661(0.037) & & &                 18.128(0.022)\nl
862.97 &18-02-98&1.3m & &      18.781(0.026) & & &                 18.277(0.019)\nl
870.97 &26-02-98&1.3m & &      18.879(0.061) & & &                 18.355(0.039)\nl
888.92 &16-03-98&1.3m & &      18.904(0.034) & & &                 18.417(0.082)\nl
 916.54&13-04-98&1.3m & &      19.260(0.063)                  \nl
1008.88&14-07-98&1.3m & &      19.630(0.067) & & &                 19.468(0.083)\nl
1026.87&01-08-98&1.3m & &      20.119(0.064) & & &                 19.714(0.051)\nl
1047.83&22-08-98&1.3m & &      20.362(0.135) & & &                 19.884(0.085)\nl
1053.83&28-08-98&1.3m & &      20.458(0.136) & & &                 19.928(0.076)\nl
1056.81&31-08-98&1.3m & &      20.472(0.101) & & &                 20.005(0.110)\nl
1062.83&06-09-98&1.3m & &      20.458(0.233)                  \nl
1072.70&16-09-98&1.3m & &      20.495(0.130) & & &                 20.149(0.116)\nl
1086.80&30-09-98&1.3m & &      20.532(0.214) \nl
1101.71&15-10-98&1.3m & &      21.246(0.269) & & &                 20.685(0.185)\nl
1107.64&21-10-98&1.3m & &      21.356(0.092) & & &                 20.970(0.234)\nl
1113.73&27-10-98&1.3m & &      21.425(0.195) \nl
\enddata
\end{deluxetable}
\clearpage

\begin{deluxetable}{cccccccc}
\scriptsize
\tablecaption{K-corrected and L$_{UVOIR}$ magnitudes for SN~1997cy.\label{tbl-3}}
\tablewidth{0pt}
\tablehead{\colhead{JD} & \colhead{B} & \colhead{V}  & \colhead{R} &  \colhead{I} & L$_{UVOIR}$ \nl
\colhead{(2450000+)}&\colhead{(mag)}&\colhead{(mag)}&\colhead{(mag)}&\colhead{(mag)}&\colhead{($\log_{10}(\rm{ergs~s}^{-1}$)) }\nl
}
\startdata
646.31 &       & 17.27  & 16.96 &&             43.32\nl
661.28 &       & 17.41  & 17.17 &&	       43.25\nl
670.21 &       & 17.48  & 17.23 &&	       43.23\nl
670.80 & 18.04 & 17.48  & 17.25 & 16.92 &           \nl
681.74 &       & 17.57  & 17.33 &&	       43.19\nl
698.79 & 18.28 & 17.72  & 17.48 &&	            \nl
700.22 &       & 17.73  & 17.49 &&	       43.12\nl
709.27 &       & 17.77  & 17.60 &&	       43.10\nl
720.16 &       & 17.87  & 17.66 &&	       43.06\nl
727.05 &       & 17.95  & 17.70 &&	       43.04\nl
739.06 &       & 18.04  & 17.77 &&	       43.01\nl
760.09 &       & 18.12  & 17.90 &&	       42.97\nl
775.06 &       & 18.27  & 18.00 &&	       42.91\nl
805.66 & 18.95 & 18.41  & 18.18 & 18.04&            \nl
814.07 &       & 18.46  & 18.37 &&	       42.80\nl
825.96 &       & 18.47  & 18.30 &&	       42.81\nl
835.04 &       & 18.57  & 18.33 &&	       42.79\nl
862.97 &       & 18.69  & 18.48 &&	       42.74\nl
870.97 &       & 18.78  & 18.56 &&	       42.70\nl
888.92 &       & 18.81  & 18.62 &&	       42.68\nl
916.54 &       & 19.17  &       &&	            \nl
1008.88&       & 19.54  & 19.67 &&	       42.33\nl
1026.87&       & 20.02  & 19.91 &&	       42.18\nl
1047.83&       & 20.27  & 20.08 &&	       42.10\nl
1053.83&       & 20.36  & 20.13 &&	       42.07\nl
1056.81&       & 20.38  & 20.21 &&	       42.05\nl
1062.83&       & 20.36  &       &&	            \nl
1072.70&       & 20.4   & 20.35 &&	       42.02\nl
1086.80&       & 20.44  &	&&                  \nl
1101.71&       & 21.15  & 20.89 &&	       41.76\nl
1107.64&       & 21.26  & 21.17 &&	       41.68\nl
1113.73&       & 21.33  &       &&	            \nl
\enddata					       
\end{deluxetable}

\end{document}